\documentclass{iopart}
\usepackage{iopams}

\begin{document}

\title[Formation of black holes]
      {Million solar mass black holes at high redshift}

\author{Oleg Y. Gnedin}

\address{Institute of Astronomy, Madingley Road, Cambridge CB3 0HA, UK}
     
\ead{ognedin@ast.cam.ac.uk}

\begin{abstract}

The existence of quasars at redshift $z > 5$ indicates that supermassive
black holes were present since the very early times.  If they grew by
accretion, the seeds of mass $\gtrsim 10^5\, M_{\odot}$ must have formed
at $z \sim 9$.  These seed black holes may result from the collapse and
dissipation of primordial gas during the early stages of galaxy
formation.  I discuss the effects of magnetic fields on the
fragmentation of cold gas clouds embedded into a hot diffuse phase and a
virialized dark matter halo.  The field of $10^{-4}$ G ejected by
supernova remnants can halt cloud break-up at $10^4\ M_{\odot}$.  High
star formation rates keep the clouds partially ionized, making ambipolar
diffusion inefficient.  The magnetically-supported clouds collapse into
black holes, which later spiral via dynamical friction into a central
cluster with the total mass $M_{bh} \sim 6\times 10^6\ M_{\odot}$.  As
the cluster collapses, the black holes merge emitting gravitational
radiation that should be detectable by LISA.

\end{abstract}

\section{Seed black holes in centers of galaxies}

Supermassive black holes have been inferred and detected in many
galaxies, distant and nearby.  They seem to be almost ubiquitous in
large elliptical galaxies (Magorrian et al. 1998).  The detection of
quasars at redshift $z>5$ (Fan et al. 1999; Stern et al. 2000) implies
that supermassive black holes existed when the Universe was only $t_H
\sim 10^9$ yr old.  This amount of time is not very large on galactic
scales and provides a powerful constraint on the formation of black
holes.

An existing black hole can grow by accretion from the gaseous disk or
disrupted stars.  Accreting with a standard efficiency $\varepsilon_M
\equiv L/\dot{M} c^2 \sim 0.1$ and emitting at a nearly Eddington rate,
$\varepsilon_L \equiv L/L_{\rm Edd} \sim 1$, the black hole doubles its
mass in
\[
t_{\rm acc} = {\varepsilon_M \over \varepsilon_L}
              {\sigma_T \, c \over 4\pi G m_p}
              \approx 4\times 10^7\ \mbox{yr}.
\]
If the accretion is inefficient ($\varepsilon_M \ll 0.1$), as in the
ADAF (Narayan \& Yi 1994) and ADIOS (Blandford \& Begelman 1999) models,
the timescale is shorter but I consider the standard scenario as an
example.  After about ten $e$-folding times, the black hole can grow in
mass by four to five orders of magnitude.  Thus, starting at $t_H
\approx 5\times 10^8$ yr (for a flat cosmological model with
$\Omega_0=0.4$ and the Hubble constant $h=0.65$) a $10^5\, M_{\odot}$
black hole in the high-redshift quasar progenitor can reach the mass
$\sim 10^9\, M_{\odot}$.

However, this efficient accretion mechanism requires a {\it seed} black
hole in place at $z \approx 9$.  At that epoch large galaxies do not yet
exist.  In the hierarchical formation scenario, they assemble via infall
and mergers of small clumps of baryonic gas and dark matter.  Even
though we may never know the exact route leading to black hole
formation, we can consider various scenarios and choose the most
plausible.  Some proposed scenarios involve stars, producing black holes
either as an endresult of the evolution of massive stars or through a
cluster of degenerate stellar remnants merging into a single object.
These processes, relying on two-body relaxation of the stellar
distribution, are slow and require a large number of local dynamical
times, at least $10^8 - 10^9$ yr (Quinlan \& Shapiro 1990; Lee 1995;
Lee, these proceedings).  Alternatively, purely gas-dynamical scenarios
seem to be more efficient (e.g., Loeb \& Rasio 1994).  The collapse and
fragmentation can proceed on the free-fall timescale, $\lesssim 10^6$ yr
at $r = 100$ pc, while the cooling time at that radius is just $5\times
10^3$ yr.

It seems that in order to form a black hole in a gas-dynamical process,
two generic constraints must be satisfied: (1) the gas must be strongly
self-gravitating; and (2) star formation must be inefficient, allowing a
large supply of cold gas to be converted to a black hole.  This second
constraint is particularly important, since star formation is a
competition to the black hole formation.

In order to prevent the formation of stars, small-scale density
perturbations should be effectively erased or damped.  One way to
achieve this is to keep the gas hot, with the thermal speed comparable
to the gravitational free-fall velocity; another is to have an almost
relativistic equation of state, where information is again transmitted
faster than the perturbations grow.  The latter criterion is satisfied
in objects dominated by radiation pressure and by isotropic
(well-tangled) magnetic field.

\section{Conservation of angular momentum}
\label{sec:gasdyn}

To set the stage, consider the just virialized dark matter halo at
$z_{\rm vir}=9$ with the mass $M_{\rm vir} \approx 1.8\times 10^{10}\,
\sigma_{100}^3\, [10/(1+z_{\rm vir})]^{3/2}\ M_{\odot}$ and the baryonic
fraction $f_b = \Omega_b/\Omega_0 = 0.1$.  Assume an isothermal density
distribution of dark matter with a velocity dispersion $\sigma_{dm} =
100\, \sigma_{100}$ km s$^{-1}$.  As the initial density perturbation
grows, the gas heats adiabatically ($T \propto \rho^{2/3}$).  During the
mergers of pregalactic clumps, the gas is shock-heated to the virial
temperature and follows the dark matter profile:
\[
   \rho_g(r) = {\sigma_{dm}^2 \over 4\pi G r^2} \, f_b(r).
\]
For the mean molecular weight of gas with the primordial composition,
$\mu_h = 0.6$, the corresponding virial temperature is $T_{\rm vir} =
3.6\times 10^5\, (1+\mu_b) \, \sigma_{100}^2$ K, where 
$\mu_b(r) \equiv {1\over r} \int f_b(r) dr$.
The initial amount of angular momentum acquired from tidal torquing at
the turn-around radius of the protogalaxy is given by the dimensionless
factor, which is remarkably constant in cosmological simulations:
\[
   \lambda = {J \, |E|^{1/2} \over G M^{5/2}} = 0.05,
\]
where $M$ is the total mass, and $E$ is the total energy.  When the gas
density in the center becomes high enough that atomic cooling is fast
($t_{\rm cool} < t_{\rm dyn}$), the gas decouples from dark matter.
Cooling gas contracts until it settles into the centrifugally-supported
disk with the characteristic size $R_d \approx \lambda \, R_{\rm vir}$.

The early epoch of a hierarchical galaxy assembly differs from the later
times in that the infalling clumps of gas and dark matter scatter off
each other, redistributing angular momentum and thus providing the
effective ``gravitational viscosity''.  Numerical simulations of
mergers, with (Mihos \& Hernquist 1996) and without (Hernquist \& Mihos
1995) the effects of star formation, show that gravitational torques
cause the gas lose a large fraction of its angular momentum, up to 99\%,
in one or two local dynamical times.  The gas in these simulations
develops strong inflows and accumulates in the center.  Any
non-axisymmetric structures, such as a temporary bar, would speed up the
process.

\section{On the possibility of a direct formation of the central object
            from gaseous collapse}
\label{sec:cenobj}

Consider now the extreme case when the gas loses most of its
angular momentum in the subgalactic mergers.
If the effective value of
$\lambda$ is reduced by a factor of hundred, $\lambda_{\rm eff} =
\lambda/5\times 10^{-4}$, the scale-length of the gaseous disk becomes a
mere 4 $\sigma_{100}$ pc.  Such disk is so dense that it may trap its
own cooling radiation and form a supermassive star.  This happens when
the speed of photon diffusion, $c/\tau$, falls below the free-fall
velocity.  Thus, the optical depth for photon scattering needs to be at
least $\tau > 210 \, \sigma_{100}^{-1}\, \lambda_{\rm eff}^{1/2}$ (a
similar scenario has been considered by Haehnelt \& Rees 1993).

There are three possible physical processes leading to photon scattering
in the primordial gas: Thomson scattering off electrons, resonant
scattering of Ly$\alpha$ photons off neutral atoms, and radiation
pressure due to recombination (Haehnelt 1995).  Because of the very high
density of the gas, the cooling time is extremely short ($\sim 8$ yr)
and therefore most of the hydrogen quickly becomes neutral.  The small
ionization fraction ($x \lesssim 10^{-5}$) renders inefficient both
electron scattering ($\tau_e \propto x$; $\tau_e \sim 0.03 \,
\sigma_{100} \, \lambda_{\rm eff}^{-1/2}$) and the radiation pressure
due to recombination ($\tau_{\rm rec} \propto x^2$; $\tau_{\rm rec} \sim
0.15 \, \sigma_{100}^{-2}$).  The dominant opacity comes from the
Ly$\alpha$ scattering ($\tau_{\rm Ly\alpha} \sim 10^5 \,
\sigma_{100}^{-2} \, \lambda_{\rm eff}^{-3/2}$).  The high density of
the gas implies a huge star formation rate, $4\times 10^4\ M_{\odot}$
yr$^{-1}$, which provides the ionizing photons that after absorption and
recombination become the Ly$\alpha$ photons.  These photons are numerous
enough to keep the supermassive object in radiation pressure support.
(Note, however, that if $\lambda > 0.01 \, \sigma_{100}^{-1/2}$ even
Ly$\alpha$ scattering would not be sufficient to trap the radiation.)

The result is a Very Massive Object, which can be either a superstar or
a very dense disk depending on the amount of angular momentum left.  The
structure of the VMO is a relativistic $\gamma=4/3$ polytrope dominated
by radiation pressure.  Such configuration is unstable to radial
collapse due to the effects of general relativity (``post-newtonian
instability'') and implodes to a black hole, if viscous torques can
remove the remaining angular momentum.  The final mass of the black hole
depends on how much mass the VMO can accumulate over its evolutionary
timescale, $\sim 3\times 10^4$ yr (Baumgarte \& Shapiro 1999).  In the
spherical collapse scenario, the gas accumulates in the center with a
constant infall rate $\dot{M} = 3.5\, \sigma_{100}^3\, M_{\odot}$
yr$^{-1}$, so the resulting black hole mass can be $\gtrsim 10^5\,
M_{\odot}$.

\section{Effects of the magnetic field}
\label{sec:mag}

If the cooling gas preserves the cosmological value of the angular
momentum, $\lambda = 0.05$, thermal instability in the disk leads to the
two-phase media of cold clouds in pressure equilibrium with hot diffuse
gas (e.g., Fall \& Rees 1985).  The dynamics of cold clouds can be
significantly modified in the presence of a magnetic field.  A
well-tangled magnetic field provides additional resistance to
gravitational contraction and prevents fragmentation of clouds below a
certain mass limit, the ``magnetic Jeans mass'' (McKee et al. 1993).
The rms magnetic energy, $\Phi_B$, would balance the gravitational pull
in clouds less massive than
\[
   M_\Phi = 0.12\, G^{-1/2}\, \Phi_B
          \approx 700 \, \left({B \over 10^{-4}\ \mbox{G}}\right)
                         \left({L \over \mbox{1 pc}}\right)^2\
                  M_{\odot},
\]
where $L$ is the size of the cloud.

There could be a primordial magnetic field in the early galaxies,
generated before virialization (Gnedin et al. 2000; Kulsrud et
al. 1997).  But additionally, after the first significant burst of star
formation, stellar winds inject metals and magnetic field lines into the
interstellar medium (Rees 1994).  The magnetic flux of each supernova
remnant would expand in space until the remnants overlap or fade away.
For example, the expanding Crab nebula is $\sim 1$ pc in size with
$B_{\rm crab} \approx 4\times 10^{-4}$ G (Hester et al. 1996).  The
number of SN remnants in the inner 100 pc would be high, $N_{SNR} \sim
3\times 10^4$, assuming a lifetime of the remnant of $10^5$ yr.  Due to
the very short mean free path in the magnetized medium, of order $10^8$
cm, the ions and electrons behave as a fluid and the shock fronts of the
SN remnants intersect but do not penetrate each other before
dissipating.  Therefore, the magnetic flux is determined locally by the
effective scale of expansion of a single SN remnant, or $\approx 3$ pc.
The average magnetic field can then be as large as $4\times 10^{-5}$ G.

The second generation of clouds formed within these supernova remnants
would be magnetized.  A large fraction of the material in the clouds is
ionized and their equilibrium temperature is therefore around $T \approx
10^4\, T_4$ K.  Such clouds can be approximated as isothermal and
uniform.  Their structure is determined by the pressure of the hot
diffuse gas.  The fraction of hot gas is determined locally as the
amount of gas unable to radiate away the heating by gravitational infall
and by other sources.  Since in the isothermal distribution, $t_{\rm
dyn} \propto r$, the density of the hot media would initially be
$\rho_h(r) \propto r^{-1}$.  After the onset of star formation from cold
gas, heating by stellar winds and supernova explosions adds to the
energy balance and couples the amounts of available hot and cold gas:
\[
   \rho_h(r) = 1.5\times 10^{-23}\ \sigma_{100}^{3/2}\ r_{100}^{-3/2}\
             \Gamma^{1/2} \left({1+\mu_b \over 3}\right)^{1/4}\
             \mbox{g cm}^{-3},
\]
with $\Gamma \equiv f_{k,0.1}\ f_{*,0.1}\ f_{b,0.1}\ \Lambda_{-23}\
\lambda_{0.05}^{-1}\ e^{-t/t_0} \sim 1$.  Here $r_{100}$ is the
galactic radius in units of 100 pc, $f_k$ is the fraction of supernova
explosion energy that goes into the kinetic energy of the surrounding
media, $f_*$ is the local fraction of cold gas that turns into stars on
a dynamical timescale, and $\Lambda = 10^{-23}\, \Lambda_{-23}$ erg
cm$^{-3}$ s$^{-1}$ is the minimum of the cooling function of the
primordial gas at $T \sim 10^6$ K.  The local supply of gas is consumed
on a timescale $t_0 = t_{\rm dyn}/f_*$ with a star formation rate
$\dot{M}_* = 43\ \sigma_{100}^3\ f_{*,0.1}\ f_{b,0.1}^{3/2}\
\lambda_{0.05}^{-3/2}\ M_\odot\ \mbox{yr}^{-1}$.  This gives the
internal cloud density
\[
  n(r) = 1.7\times 10^3\ \sigma_{100}^{7/2}\ r_{100}^{-3/2}\
                 T_4^{-1}\ \Gamma^{1/2}
                 \left({1+\mu_b \over 3}\right)^{5/4}\ \mbox{cm}^{-3},
\]
and the Jeans mass
\[
   M_J \approx 3.4\times 10^5\ \sigma_{100}^{7/4}\ r_{100}^{3/4}\
               T_4^2\ \Gamma^{-1/4}\
               \left({1+\mu_b \over 3}\right)^{-5/8}\ M_{\odot}.
\]

If the cloud cools enough that the value of the Jeans mass falls below
its mass, $M_{cl}$, the cloud starts contracting and fragmenting into
smaller, denser subclouds which cool further.  This process can go on
until $M_{cl} \sim M_\Phi$, at which point the magnetic pressure will
resist fragmentation.  Since both the gravitational and magnetic energies
scale with the cloud size as $L^{-1}$, the cloud will remain in marginal
equilibrium at any value of $L$.  Thus, any increase in the external
pressure will cause uniform cloud collapse.  Unless the magnetic flux
can be effectively dissipated, the inevitable outcome of such
contraction is the formation of a compact supermassive object.

In Galactic molecular clouds, the magnetic flux is removed by ambipolar
diffusion of ions through neutral atoms.  The diffusion
timescale is
\[
   t_{\rm ad} \approx 9\times 10^{11} \, x
                \left({L \over 1\, \mbox{pc}}\right)^2
                \left({n \over 10^3\, \mbox{cm}^{-3}}\right)^2
                \left({B \over 10^{-4}\, \mbox{G}}\right)^{-2}
                \mbox{yr},
\]
where $L$ is the scale of the magnetic field variation (Draine, Roberge
\& Dalgarno 1983).  The efficiency of ambipolar diffusion depends on the
ionization fraction, $x$.  The cores of molecular clouds are shielded
from external ionizing radiation ($x \lesssim 10^{-7}$) and the magnetic
field decays to allow further fragmentation and star formation.  In the
primordial clouds, the X-rays from first star clusters should be able to
penetrate the whole cloud.  The X-ray luminosity at 1 keV resulting from
a solar mass of star formation per year can be empirically estimated to
be $\sim 3\times 10^{40}$ erg s$^{-1}$ (Oh 2001).  This leads to a
significant ionization fraction, $x_c \sim 2\times 10^{-3}$, and a long
ambipolar diffusion time, $t_{\rm ad} \sim 2\times 10^9$ yr.  Therefore,
the clouds with mass $M_\Phi$ would remain magnetically supported.
Also, a frozen uniform magnetic field is important for removing cloud's
angular momentum via ``magnetic braking'' on roughly a dynamical time of
the ambient medium (McKee et al. 1993).

The dynamics of cold clouds is regulated by two competing effects,
condensation from the hot phase and destruction in predominantly
inelastic collisions.  Thermal instability of the virialized gas leads
to the formation of clouds on a cooling timescale.  The binding energy
of these clouds is relatively small; the escape velocity $\sim 2\,
(M_3/L_{pc})^{1/2}$ km s$^{-1}$.  The clouds in the disk move with a
much higher velocity, $v_{\rm rel} \gtrsim 10$ km s$^{-1}$, and
therefore cloud collisions would quickly disband them.  The time between
collisions depends on the number of clouds within the inner 100 pc.
Assuming that the clouds develop a scale-free mass distribution
$dN_{cl}/dM \propto M^{-2}$ with $M_{min} \approx M_\Phi$ and $M_{max}
\approx M_J$, the smallest clouds would survive long enough for magnetic
braking to operate if $M_\Phi \approx 3\times 10^4\, r_{100}^{-3/4} \,
\sigma_{100}^{-13/4}\, M_{\odot}$.

Once the magnetically-supported cloud loses enough angular momentum,
external pressure perturbations would cause an unlimited uniform
collapse to a black hole through an intermediate stage of a supermassive
star.  During the radiation-supported VMO phase, strong winds and
outflows may reduce the final mass by roughly a half, although the exact
value is uncertain.  The VMOs close to the center would spiral in via
dynamical friction off the gas and stars.  Allowing time $t_{\rm df} =
3\times 10^8$ yr for the dynamical friction with the circular velocity
$\sim 170\ \sigma_{100}$ km s$^{-1}$, the radius of the swept-up region
is about $r_{\rm df} = 30$ pc (Binney \& Tremaine 1987) and the critical
cloud mass $M_\Phi = 1.5\times 10^4\ \sigma_{100}^{-7}\ M_{\odot}$.
This requires the magnetic field $B \sim 4\times 10^{-4}$ G on the scale
of the cloud, $L_\Phi \sim 2$ pc.  The number of $M_\Phi$ clouds within
the central 30 pc is $\sim 780$, which including the mass loss of the
VMO results in the cluster of black holes with the total mass $\sim
6\times 10^6\ M_{\odot}$.  The cluster keeps contracting until it
dominates the gravity in the center and the dynamical friction becomes
inefficient.  The size of the black hole cluster at that point is
$R_{cl} = 0.9\ \sigma_{100}^{-2}\ \lambda_{0.05}$ pc.

Notice the strong dependence of the black hole mass on the value of the
magnetic field and the circular velocity of the halo, $M_{bh} \propto
M_\Phi \propto B^3 \, \sigma^{-7}$.  Small galaxies forming first would
have a higher mass of the magnetized clouds and also require a weaker
magnetic field to support them.

\section{Gravitational radiation from merging black holes}

The subsequent evolution of the cluster of black holes is relatively
well understood.  A self-similar collapse occurs (Cohn 1980) until
binaries form via gravitational radiation (Quinlan \& Shapiro 1987),
their eqs. (41)-(44).  Because of the small number of objects in the
cluster the initial relaxation time is short, $t_{\rm rel} = 6.5\times
10^4$ yr, but the initial binary capture time is long, $t_{\rm cap} =
3.3\times 10^{10}$ yr.  Binary formation becomes effective when the
cluster core collapses to $\sim 0.02\ R_{cl}$ on the collisionless
timescale $t_{cc} = 330\, t_{\rm rel} \sim 2\times 10^7$ yr.  Since this
time is shorter than the local Hubble time ($t_H = 5\times 10^8$ yr),
most of the black holes undergo hierarchical binary mergers through
gravitational wave radiation.

The merger of a large number of black holes would lead to a significant
amount of gravitational radiation from the galactic center.  The
comoving density of halos with $M_{\rm vir} = 2\times 10^{10}\,
M_{\odot}$ at $z=9$ is $\sim 7\times 10^{-3}$ Mpc$^{-3}$.  A coalescence
of two $10^4\ M_{\odot}$ black holes would lead to a gravitational wave
amplitude $h \sim 0.1\, r_g/R_H \sim 3\times 10^{-20}\, M_4$, where
$r_g$ is the gravitational radius of the black hole and $R_H$ is the
Hubble distance.  The maximum frequency of gravitational radiation is
$\nu_{\rm max} \lesssim 0.1\, c/r_g \sim 1\ M_4^{-1}$ Hz.  The mergers
would proceed hierarchically, doubling the mass in each event.  Thus,
the last merger of the $3\times 10^6\, M_{\odot}$ black holes gives the
maximum amplitude $h \sim 10^{-17}$.

Such gravitational signal should be detectable by LISA.  The early black
holes should originate in the highest density peaks and therefore be
highly clustered.  They should appear in the same direction as the
high-redshift quasars.

\bigskip 
\noindent
I would like to thank M. Rees, M. Begelman, J. P. Ostriker, and
E. Zweibel for critical discussions, and the LISA conference organizers
for hospitality.

\section*{References}
\begin{harvard}

\item[]
Baumgarte T. W., Shapiro S. L., 1999, ApJ, 526, 941

\item[]
Binney J., Tremaine S., 1987, Galactic Dynamics 
  (Princeton: Princeton University Press)

\item[]
Blandford R. D., Begelman M. C., 1999, MNRAS, 303, L1

\item[]
Cohn H., 1980, ApJ, 242, 765

\item[]
Draine B. T., Roberge W. G., Dalgarno A., 1983, ApJ, 264, 485

\item[]
Fall S. M., Rees M. J., 1985, ApJ, 298, 18

\item[]
Fan X. et al. (SDSS collaboration), 1999, AJ, 118, 1

\item[]
Gnedin N. Y., Ferrara A., Zweibel E. G., 2000, ApJ, 539, 505

\item[]
Haehnelt M. G., 1995, MNRAS, 273, 249

\item[]
Haehnelt M. G., Rees M. J., 1993, MNRAS, 263, 168

\item[]
Hernquist L., Mihos J. C., 1995, ApJ, 448, 41

\item[]
Hester, J. J. et al., 1996, ApJ, 456, 225

\item[]
Kulsrud R. M., Cen R., Ostriker J. P., Ryu D., 1997, ApJ, 480, 481

\item[]
Lee H. M., 1995, MNRAS, 272, 605

\item[]
Loeb A., Rasio F. A., 1994, ApJ, 432, 52

\item[]
Magorrian J. et al., 1998, AJ, 115, 2285

\item[]
McKee C. F., Zweibel E. G., Goodman A. A., Heiles C., 1993,
  in Protostars and Planets III, ed. E. H. Levy \& J. I. Lunine
  (Tucson: Univ. Arizona Press), 327

\item[]
Mihos J. C., Hernquist L., 1996, ApJ, 464, 641

\item[]
Narayan R., Yi I., 1994, ApJ, 428, L13

\item[]
Oh S. P., 2001, ApJ, 553, 499

\item[]
Quinlan G. D., Shapiro S. L., 1987, ApJ, 321, 199

\item[]
Quinlan G. D., Shapiro S. L., 1990, ApJ, 356, 483

\item[]
Rees M. J., 1994, in Cosmical Magnetism (Kluwer), ed. D. Lynden-Bell, 155

\item[]
Stern D. et al., 2000, ApJ, 533, L75

\end{harvard}
\end{document}